\documentclass[structabstract]{aa}
\usepackage{natbib}
\usepackage{hyperref}
\usepackage{amsmath}
\hypersetup{colorlinks=true,citecolor=blue}
\usepackage{graphicx}
\usepackage{draftcopy}
\usepackage{url}
\usepackage{amsmath}
\usepackage{amssymb} 
\usepackage{amsfonts}
\usepackage[ruled]{algorithm2e}
\usepackage{graphicx,epsfig,graphics}
\usepackage{aas_macros}
\usepackage{graphics}
\usepackage{algorithmic}
\usepackage{stmaryrd}
\usepackage{ulem}
\usepackage{caption}
\usepackage{subcaption}

\DeclareMathOperator{\prox}{prox}
\DeclareMathOperator{\argmin}{\operatornamewithlimits{arg\,min}}

\title{PRISM: Sparse Recovery of the Primordial Power Spectrum}
\author{ P. Paykari \inst{1} \thanks{paniez.paykari@cea.fr} \and F. Lanusse \inst{1} \thanks{francois.lanusse@cea.fr} \and  J.-L. Starck \inst{1} \and  F. Sureau \inst{1}    \and  J. Bobin \inst{1}}
\institute{$^1$ Laboratoire AIM, UMR CEA-CNRS-Paris 7, Irfu, SAp, CEA Saclay, F-91191 Gif sur Yvette cedex, France.}

\begin{document}
\label{firstpage}
\date{\today}

\abstract{{
} }
{
The primordial power spectrum describes the initial perturbations
in the Universe which eventually grew into the large-scale structure
we observe today, and thereby provides an indirect probe of inflation
or other structure-formation mechanisms. Here, we introduce a new method
to estimate this spectrum from the empirical power spectrum of cosmic microwave
background (CMB) maps.
}
{
A sparsity-based linear inversion method, coined \textbf{PRISM}, is presented. This technique leverages a sparsity prior on features in the primordial power spectrum in a wavelet basis to regularise the inverse problem. This non-parametric approach does not assume a strong prior on the shape of the primordial power spectrum, yet is able to correctly reconstruct its global shape as well as localised features. These advantages make this method robust for detecting deviations from the currently favoured scale-invariant spectrum.
}
{
We investigate the strength of this method on a set of WMAP 9-year simulated data for three types of primordial power spectra: a nearly scale-invariant spectrum, a spectrum with a small running of the spectral index, and a spectrum with a localised feature. This technique proves to easily detect deviations from a pure scale-invariant power spectrum and is suitable for distinguishing between simple models of the inflation. We process the WMAP 9-year data and find no significant departure from a nearly scale-invariant power spectrum with the spectral index $n_s = 0.972$.
}
{
A high resolution primordial power spectrum can be reconstructed with this technique, where any strong local deviations or small global deviations from a pure scale-invariant spectrum can easily be detected. 
}

\maketitle

\section{Introduction}
\label{sec:intro}

The primordial power spectrum encodes the physics of the early Universe and its measurement 
is one of the key research areas in modern cosmology. Amongst the proposed models which describe the early Universe, inflation \citep{guth,Linde-inflation1982} 
is currently the most favoured one. In this model early perturbations are produced by quantum fluctuations during the epoch of an accelerated expansion. These perturbations then grow into the large scale structure we observe today. The simplest models of inflation predict almost purely adiabatic primordial perturbations with a 
nearly scale-invariant power spectrum. In these models the primordial power spectrum is often described in terms of a spectral index $n_s$ and an amplitude of the perturbations $A_s$ as
\begin{equation}
P(k)=A_s \left(\frac{k}{k_p}\right)^{n_s-1}\;,
\label{eq:pkdef}
\end{equation}
where $k_p$ is a pivot scale. This spectrum represents the initial conditions
set at inflation. The simplest ansatz for characterising the primordial perturbations is the so-called Harrison-Zeldovich (HZ) model, which sets $n_s=1$ \citep{Harrison1970,Zeldovich1972}. This is an exact scale-invariant spectrum, which has been ruled out by different datasets, as will be discussed later. Instead, the near scale-invariant spectrum with $n_s<1$ fits the current observations very well \citep[for e.g.,][]{PlanckCP}. 
However, numerous models have been proposed for the generation of the perturbations, predicting deviations from the perfectly scale-invariant power spectrum.
The simplest are the slow-roll inflationary models which describe the deviations through 
a minimal scale dependence of the spectral index of the power spectrum, the so-called `running' $\alpha_s$, formulated as 
\begin{equation}
P(k)=A_s \left(\frac{k}{k_p}\right)^{n_s-1+\frac{1}{2}\alpha_s\ln \left({k}/{k_p}\right)}\;.
\label{eq:pkdef}
\end{equation}
More complex 
models generating deviations from scale-invariance include those with features on the potential \citep{Starobinsky:1992ts,Adams:2001vc,Wang:2002hf,Hunt:2004vt,Joy:2007na,Hunt:2007dn,Pahud:2008ae,Lerner:2008ad,Kumazaki2011,Meerburg2012},
a small number of $e$-folds \citep{Contaldi:2003zv,Powell:2006yg,Nicholson:2007by},
or other exotic inflationary models \citep{Lesgourgues:1999uc,Feng:2003zua,Mathews:2004vu,Jain:2008dw,Romano:2008rr}. Therefore, determining the shape of the primordial power spectrum will allow us to evaluate how well these models of the early Universe compare to the observations, rule out some of the proposed models, and thus giving us a better intuition into the conditions of the primordial Universe.

A few probes of the physics of the early Universe include non-Gaussianity, the primordial tensor power spectrum,
a cosmic gravitational wave background and a cosmic neutrino background,
none of which have been observed with an acceptable significance. 
On the other hand, we can observe $P(k)$ through the windows of the Cosmic Microwave Background (CMB) and Large Scale Structure (LSS), which are incredibly important and powerful insights into
the early Universe.

The recent Planck mission temperature anisotropy data, combined
with the WMAP large-angle polarisation, constrain the scalar spectral index to $n_{s}=0.9603 \pm0.0073$ \citep{PlanckCP}, which rules out exact scale-invariance at over $5 \sigma$. In addition, Planck does not find a statistically significant running of the scalar spectral index, obtaining $\alpha_s =-0.0134\pm0.0090$. In \citet{PlanckCP} an extensive investigation is performed by the Planck collaboration to see whether the primordial power spectrum contains any features. They report that a penalised likelihood approach suggests a feature
near the highest wavenumbers probed by Planck at an estimated significance of $\sim 3 \sigma$. In addition, a parameterised oscillatory feature does improve the fit to the data by $\Delta\chi^2_{\mathrm{eff}} \approx 10$, however Bayesian evidence does not prefer these models. On the other hand, high resolution CMB experiments, such as the South Pole telescope (SPT)\footnote{\url{http://pole.uchicago.edu/spt/index.php}}, detect a small running of the spectral index; $-0.046 < \alpha_s < -0.003$ at $95\%$ confidence \citep{SPT_cos}. In general, any detections of the running of the spectral index have been small and consistent with zero. Therefore, a highly sensitive algorithm is required to detect these small deviations.

There are generally two approaches to determine the shape of the primordial power spectrum, one is by parametrisation and the second is a reconstruction. Numerous parametric approaches search for features with a similar form to those in complex inflationary models have been performed along with a simple
binning of $P(k)$ \citep{Bridle:2003sa,Contaldi:2003zv,Parkinson:2004yx,Sinha:2005mn,Sealfon:2005em,Mukherjee:2005dc,Bridges:2005br,Bridges:2006zm,Covi:2006ci,Joy:2008qd,Verde:2008zz,PaykJaffe_OB,GuoEtAl2011,GoswamiPrasad2013}. 
Non-parametric methods, which make no assumptions about the model of the early
Universe, have also been probed \citep{Hannestad:2000pm,Wang:2000js,Matsumiya:2001xj,Shafieloo:2003gf,Bridle:2003sa,Kogo:2003yb,Mukherjee:2003cz,Mukherjee:2003ag,Hannestad:2003zs,Kogo:2004vt,TocchiniValentini:2004ht,Leach:2005av,Shafieloo:2006hs,Shafieloo:2007tk,Nagata:2008tk,Nagata:2008zj,Nicholson:2009pi,NichPaykCont,HazraEtAl20131}. 
For an extensive review on how to search for features in the primordial power spectrum using a wide range of methods, refer to the following papers and the references therein, which provide a sample on non-parametric reconstruction:  deconvolution
\citep{Tocchini-Valentini2006,IchikiNagata2009,IchikiEtAl2010} (including Richardson-Lucy deconvolution \citep{Lucy, Richardson,Hamann2010, Shafieloo:2007tk}), smoothing splines \citep{VP-mini-param-pPS,PeirisVerde2010, Sealfon:2005em,GauthierBucher2012}, linear interpolation \citep{Hannestad:2003zs,Bridle:2003sa}, and Bayesian model selection \citep{BridgesEtAl2009,VazquezEtAl2012}. 

Non-parametric methods are hampered by the non-invertibility of the transfer function that descries the {\it transfer} from $P(k)$ to CMB (or LSS). Specifically for the CMB power spectrum, the dependence on the transfer function has the form
\begin{equation} 
C_{\ell}^{\textrm{th}}=4\pi\int_{0}^{\infty}d\ln k\Delta_{\ell}^{2}(k)P(k)\;,
\label{eq:CMB}
\end{equation} 
where $\ell$ is the angular wavenumber that corresponds to an angular scale via $\ell\sim180^{o}/\theta$
and $\Delta_{\ell}(k)$ is the angular transfer function of the radiation anisotropies, which holds the cosmological parameters responsible for the evolution of the Universe. 
As the CMB spectrum is \textit{jointly} sensitive to the primordial spectrum and the cosmological parameters in the transfer function, there is an induced degeneracy between them. The impact and level of this degeneracy have been investigated in \citep{PaykJaffe_OB}. A joint estimation of the cosmological parameters and a free form primordial power spectrum would be prohibitively expensive to perform (as the parameter space potentially becomes very large). As a result, a parametric form of the primordial power spectrum is assumed when jointly estimating this spectrum along with the other cosmological parameters. This hides any degeneracies between the cosmological parameters in the transfer function and the form of $P(k)$. Thus it is not clear what the significance of any
features found in the reconstructed $P(k)$ should be. One way to break this induced degeneracy is by adding extra information, such as polarisation or LSS data \citep{Hu:2003vp,Nicholson:2009pi,Mortonson:2009qv}.

The other hurdle into the estimation of the primordial spectrum is that this continuous spectrum is deconvolved from discrete data $C_{\ell}$. This causes problems if the primordial power spectrum contains features that are smaller or comparable to the gridding in $\ell$ ($\Delta \ell=1$). This limits our ability to fully recover the 
primordial power spectrum; in the case of the CMB, even a perfect survey cannot recover the primordial power spectrum completely \citep{ho}. 

Here, we propose a new non-parametric method for the reconstruction of the primordial power spectrum from CMB data, which is based on the sparsity of the primordial power spectra in a wavelet basis and an appropriate noise modelling of the CMB power spectrum \citep{Tousi}. 


\subsection*{Paper content}
\label{sec:content}

In section~\ref{sec:method} we present the primordial power spectrum reconstruction problem and describe the technique we have developed to perform the reconstruction. Our algorithm is tested on three sets of simulated spectra and applied to WMAP 9-year data in section~\ref{sec:results}. In section~\ref{sec:conclusion} we conclude and state some potential perspectives.

\section{Sparse Recovery of the Primordial Power Spectrum}
\label{sec:method}

\subsection{Empirical power spectrum}

A CMB experiment, such as Planck, measures the CMB temperature anisotropy $\Theta(\vec{p})$ in direction $\vec{p}$, which is described as $T(\vec{p}) = T_{\mathrm{CMB}} [1 + \Theta(\vec{p}) ]$. 
This anisotropy field can be expanded in terms of spherical harmonic functions $Y_{\ell m}$ as
\begin{equation}
\Theta(\vec{p}) = \sum\limits_{\ell=0}^{\infty} \sum\limits_{m=-\ell}^{\ell} a_{\ell m} Y_{\ell m}(\vec{p}) \;,
\end{equation}
with $a_{\ell m}$ being the spherical harmonic coefficients. The CMB anisotropy $\Theta(\vec{p})$ is assumed to be Gaussian distributed, which makes the $a_{\ell m}$ independent and identically distributed (i.i.d.) Gaussian variables with zero mean, $\langle a_{\ell m} \rangle = 0$, and variance
\begin{equation}
\langle a_{\ell m} a^*_{\ell^{\prime} m^{\prime}} \rangle = \delta_{\ell \ell^{\prime}} \delta_{mm^{\prime}} C_\ell^{\textrm{th}} \;,
\label{eq:ps}
\end{equation}	
where $C_\ell^{\textrm{th}}$ is the CMB temperature angular power spectrum introduced in Equation~\ref{eq:CMB}.
However, we only observe a realisation of this underlying power spectrum on our sky, which we can estimate using the \textit{empirical power spectrum estimator} defined as
\begin{equation}
\widehat{C}^{\mathrm{th}}_\ell = \frac{1}{2\ell + 1} \sum\limits_{m=-\ell}^{\ell} | a_{\ell m} |^2\;,
\label{eq:empirical-ps}
\end{equation}
where $\widehat{C}^{\mathrm{th}}_\ell$ is an unbiased estimator of the true underlying power spectrum; $\langle \widehat{C}^{\mathrm{th}}_\ell\rangle = C_\ell^{\textrm{th}}$, in the case of noiseless CMB data over full sky.

For a given $\ell$, the empirical power spectrum follows a $\chi^2$ distribution with $2\ell +1$ degrees of freedom, as it is a sum of the squares of independent Gaussian random variables. To account for this variability, we recast the relation between $\widehat{C}^{\mathrm{th}}_\ell$ and $C^{\mathrm{th}}_{\ell}$ as
\begin{equation}
\widehat{C}^{\mathrm{th}}_\ell = C^{\mathrm{th}}_\ell \; Z_\ell \;,
\label{eq:multnoise}
\end{equation} 
where $Z_\ell= \sum_{m} | a_{\ell m} |^2/LC^{\mathrm{th}}_\ell$, which is a random variable representing a multiplicative noise distributed according to:
\begin{equation}
\label{eq:chi2noise}
L Z_{\ell} \sim \chi^2_{L} \qquad \mbox{ where $L = 2 \ell + 1$} \;.
\end{equation} 

In particular, the standard deviation of the empirical power spectrum estimator for a given $\ell$ is $\sqrt{\left(2/L\right)} \; C^{\mathrm{th}}_\ell$.

%
%
%
%
%
%
%

\subsection{Accounting for instrumental noise and partial sky coverage}

So far, we have considered that the CMB anisotropy data was available on the full sky which is not possible in practice due to the different Galactic foregrounds. Applying a mask on the sky results in the following modification of the spherical harmonic coefficients of the CMB temperature anisotropy:
\begin{equation}
	\tilde{a}_{\ell m} = \int \Theta(\vec{p}) W(\vec{p}) Y_{\ell m}^*(\vec{p}) d\vec{p}\;,
\end{equation}
where $W(\vec{p})$ is the window function applied to the data. The presence of the window function induces correlations between the $a_{\ell m}$ coefficients at different $\ell$ and different $m$ and hence Equation~\ref{eq:ps} is no longer true.

One can define the \textit{pseudo power spectrum} $\widetilde{C}_{\ell}$ as the application of the empirical power spectrum estimator on the spherical harmonic coefficients of the masked sky. In case of data contaminated with additive Gaussian stationary noise, the pseudo power spectrum is
\begin{equation}
	\widetilde{C}_\ell = \frac{1}{ 2 \ell + 1 } \sum_{m = -\ell}^{\ell} | \tilde{a}_{\ell m} + \tilde{n}_{\ell m} |^2 \;,
\end{equation}
where $\tilde{n}_{\ell m}$ are the spherical harmonic coefficients of the masked instrumental noise. 

Following the MASTER method from \citet{MASTER}, the pseudo power spectrum $\widetilde{C}_\ell$ and the empirical power spectrum $\widehat{C}^{\mathrm{th}}_\ell$ can be related through their ensemble averages:
\begin{equation}
\langle \widetilde{C}_\ell \rangle = \sum_{\ell^\prime} M_{\ell \ell^\prime}   \langle \widehat{C}^{\mathrm{th}}_{\ell^\prime} \rangle +  \langle \widetilde{N}_\ell \rangle \;,
\end{equation}
where $M_{\ell \ell^\prime}$ describes the mode-mode coupling between modes $\ell$ and $\ell^\prime$ resulting from computing the transform on the masked sky. Note that in this expression, $ \langle \widehat{C}^{\mathrm{th}}_{\ell^\prime} \rangle =  C^{\mathrm{th}}_{\ell^\prime} $ and we introduce the following notations:
\begin{equation}
C_\ell = \langle \widetilde{C}_\ell \rangle \qquad \mbox{ and } \qquad N_\ell = \langle \widetilde{N}_\ell \rangle \;.
\end{equation}
Please note that $C_\ell$ and $N_\ell$ refer to the CMB and the noise power spectra of the masked maps respectively.

We will further work under the approximation that the pseudo power spectrum $\widetilde{C}_\ell$ still follows a $\chi^2$ distribution with $2 \ell + 1$ degrees of freedom and can be modelled as:
\begin{align}
	\widetilde{C}_\ell &= C_\ell Z_\ell \;, \\
	 &=	 \left( \sum_{\ell^\prime} M_{\ell \ell^\prime} C^{\mathrm{th}}_{\ell^\prime} +   N_\ell \right) Z_\ell \label{eq:pseudoToTheoPS} \;,
\end{align}
where $Z_\ell$ is defined in Equation~\ref{eq:chi2noise}.

\subsection{Formulation of the inverse problem}

Now we aim to estimate the primordial power spectrum $P_{k}$ from the pseudo power spectrum $\widetilde{C}_\ell$ computed on a masked noisy map of the sky.

Equation~\ref{eq:pseudoToTheoPS} relates the observables $\widetilde{C}_{\ell}$ to the theoretical CMB anisotropy power spectrum $C_\ell^{\mathrm{th}}$, taking into account instrumental noise, sample variance and masking. $C_\ell^{\textrm{th}}$ is itself related to the primordial power spectrum through the convolution operation defined in Equation~\ref{eq:CMB}. For a finite sampling of the wavenumber $k$, this convolution can be recast as a matrix operator $\mathbf{T}$ acting on the discretely sampled primordial spectrum, now referred to as $P_{k}$,
\begin{equation}
C_{\ell}^{\textrm{th}} \simeq \sum\limits _{k}T_{\ell k}P_{k}\;,
\label{eq:rld-start}
\end{equation}
with matrix elements $T_{\ell k}=4 \pi \Delta\ln k\,\Delta_{\ell k}^2$, where $\Delta\ln k$ is the logarithmic $k$ interval for the discrete sampling chosen in the integration of the system of equations. Due to the non-invertibility of the $\mathbf{T}$ operator, recovering the primordial power spectrum $P_k$ from the true CMB power spectrum $C_\ell^{\textrm{th}}$ constitutes an ill-posed inverse problem.
Finally, the complete problem we aim to solve can be condensed in the following form:
\begin{equation}
\widetilde{C}_\ell = \left( \sum_{\ell^\prime k } M_{\ell \ell^\prime} T_{\ell^\prime k}P_{k} +   N_\ell \right) Z_\ell\;.
\label{eq:complete-problem-mult}
\end{equation}
We assume that the masked instrumental noise power spectrum $N_\ell$ is known for a given experiment. It can be computed from a JackKnife data map or from realistic instrumental noise simulations. Therefore, in the power spectrum of the data $\widetilde{C}_\ell$, only the primordial power spectrum $P_k$ remains unknown. Note that we assume that cosmology is known and hence operator $\mathbf{T}$ is known. 

The presence of the multiplicative noise $Z_\ell$ further complicates the ill-posed inverse problem of Equation~\ref{eq:rld-start}. We address both the inversion problem and the control of the noise in the framework of sparse recovery. Indeed, the inversion problem in Equation~\ref{eq:complete-problem-mult} can be regularised in a robust way by using the sparse nature of the reconstructed signal as a prior. Furthermore, sparse recovery has already been successfully used in the TOUSI algorithm \citep{Tousi} to handle the multiplicative noise term and denoise the CMB power spectrum with high accuracy from single realisations.

\subsection{The TOUSI method}
\label{subsec:tousi}

It was shown in \citet{Tousi} that the a theoretical power spectrum $C_{\ell}^{\textrm{th}}$ can be represented with only a few coefficients (i.e. sparse representation) in a given dictionary (e.g., wavelet, DCT, etc.) and that a sparse regularisation allows us to recover the theoretical power spectrum directly from the measured CMB empirical power spectrum $\widehat{C}^{\mathrm{th}}_{\ell}$, without having to know the cosmological parameters.

A proper treatment of the non-Gaussian noise on $\widehat{C}^{\mathrm{th}}_{\ell}$ was proposed in TOUSI, which is based on the Wahba variance stabilisation transform (VST). After the variance stabilisation is applied, the noise on $\widehat{C}^{\mathrm{th}}_{\ell}$ can be treated as an additive Gaussian noise with zero mean and unit variance. The VST operator $\mathcal{T}$ is defined as
\begin{equation}
\mathcal{T} : x \in \mathbb{R}^+ \mapsto \frac{ \ln x - \mu_L}{\sigma_L} \;,
\label{eq:vst}
\end{equation}
where $\mu_L = \psi_0(L/2) - \ln (L/2)$ and $\sigma_L^2 = \psi_1(L/2)$, where $\psi_m$ is the polygamma function $\psi_m(t) = \frac{d^{m+1}}{dt^{m+1}} \ln \Gamma(t)$. We note $C^s_\ell$ as the stabilised empirical power spectrum after applying the VST and get
\begin{equation}
C^s_\ell = \mathcal{T}(\widehat{C}_\ell^{\mathrm{th}} ) = \frac{\ln C^{\textrm{th}}_\ell}{\sigma_L} + \epsilon_\ell \;,
\label{eq:VST_Cl}
\end{equation} 
where $ \epsilon_\ell = (\ln Z_\ell - \mu_L)/\sigma_L \sim \mathcal{N}(0,1)$. We define the inverse operator of $\mathcal{T}$ as 
\begin{equation}
\mathcal{R}: x \in \mathbb{R} \mapsto \exp(\sigma_L x) \;.
\end{equation}
Having $X_\ell$ as the unknown power spectrum to be recovered, the TOUSI method consists in minimising the following constrained optimisation problem:
\begin{equation}
\label{eq_min_supp}
\min_{X_\ell} \| { \mathbf{\Phi}}^{t}{X_\ell}\|_1 \quad \mathrm{s.t.} \quad 
\begin{cases}  
X_\ell \geqslant 0 \\  
S  \odot \big(\mathbf{\Phi}^{t}{\cal T}(Y_\ell)\big) = S  \odot \big(\mathbf{\Phi}^{t} C^s_\ell \big)
\end{cases},
\end{equation}
where $Y_\ell = X_\ell+N_\ell^{\mathrm{th}}$, $\odot$ stands for the Hadamard product (i.e. entry-wise multiplication) of two vectors and $\mathbf{\Phi}$ is the chosen dictionary. Vector $S$ provides a set of active coefficients (not due to noise), where $S_{i} = 1$ if the $i$th coefficient 
 $\left(\mathbf{\Phi}^t {\cal T}(Y_\ell)\right)_i$ is above the noise level (i.e. significant) and 0 otherwise. 
This minimisation is performed iteratively:
\begin{equation}
\begin{split}
\widetilde{X}_\ell &= {\cal R} \left({\cal T} \left(Y_\ell^n\right)+{\mathbf{\Phi}} S \odot \left({\mathbf{\Phi}}^{t} \left(C_\ell^{s}-{\cal T} \left(Y_\ell^n\right)   \right)   \right)\right) - N_\ell^{\mathrm{th}}\;,\\
X_\ell^{n+1} &= \mathcal{P}_{+}\left(  {\mathbf{\Phi}} ~ \text{ST}_{\lambda_n}({ \mathbf{\Phi}}^{t}\widetilde{X}_\ell) \right) \;,
\label{eq:eq2}
\end{split}
\end{equation}
where $n$ is the iteration number, $\mathcal{P}_{+}$ is a positivity constraint. The soft thresholding operator $\mathrm{ST}_{\lambda_n}$ has an iteration dependent threshold level $\lambda_n$ and is defined as
\begin{equation}
\forall \mathbf{x} \in \mathbb{R}^n, \ \mathrm{ST}_\lambda(\mathbf{x})_i = sgn(x_i) ( | x_i | - \lambda )_{+}.
\end{equation}

Full details of the TOUSI algorithm can be found in \citet{Tousi}.

\subsection{$P{k}$ sparse recovery formulation}

The problem of reconstructing the primordial power spectrum is stated in Equation~\ref{eq:complete-problem-mult}. Solving this problem has three inherent difficulties:
1. the singularity of the convolution operator $T_{\ell k}$, which makes the inverse problem ill-posed even in the absence of noise;
2. the multiplicative noise on the power spectrum;
3. the mask applied to the maps, inducing correlations on the power spectrum.

To address the inverse problem, we adopt the sparse regularisation framework. If the signal to recover (in our case $P_k$) can be sparsely represented in an adapted dictionary $\mathbf{\Phi}$ then this problem, known as the `basis pursuit denoising' BPDN, can be recast as an optimisation problem. In the case of the inverse problem stated in Equation~\ref{eq:complete-problem-mult}, the optimisation problem can be formulated as:
\begin{equation}
\min\limits_X \frac{1}{2} \parallel C_\ell - (\mathbf{M}\mathbf{T} X + N_\ell) \parallel_2^2 + \lambda \parallel \mathbf{\Phi}^t X \parallel_0\;,
\label{eq:bp-lagragian}
\end{equation}
where $X$ is the reconstructed estimate for the primordial power spectrum $P_k$. The first term in equation (\ref{eq:bp-lagragian}) imposes a $\ell_2$ fidelity constraint to the data while the second term promotes the sparsity of the solution in dictionary $\mathbf{\Phi}$. The parameter $\lambda$ tunes the sparsity constraint.

One can notice that in Equation~\ref{eq:bp-lagragian}, only the ensemble mean of the pseudo power spectrum $ C_\ell $ appears (which is unknown) and not the actual measurements $\widetilde{C}_\ell$. This is linked to the second difficulty, the measurements are contaminated with a multiplicative noise which cannot be handled with the formulation of Equation~\ref{eq:bp-lagragian}. Indeed this formalism holds for measurements contaminated with additive Gaussian noise which is not the case of the $\widetilde{C}_\ell$. To overcome this issue, we use the variance stabilisation introduced in the TOUSI algorithm. 

Let $R_\ell(X)$ be the residual between $C_\ell$ and the reconstructed CMB power spectrum given a primordial power spectrum $X$, $C_\ell(X) = (\mathbf{M}\mathbf{T} X + N_\ell)$:
\begin{equation}
	R_\ell(X) = C_\ell - C_\ell(X)\;.
\end{equation}
Note that $R_\ell(X)$ is the data fidelity term in Equation~\ref{eq:bp-lagragian}. Since $C_\ell$ is unknown, so is $R_\ell(X)$, but we can estimate it from the data $\widetilde{C}_\ell$. Let us consider the following difference:
\begin{align}
\mathcal{T}(\widetilde{C}_\ell) - \frac{\ln( C_\ell(X) )}{\sigma_L}  &= \frac{ \ln(C_\ell) - \ln( C_\ell(X) ) }{ \sigma_L}  + \epsilon_\ell\;, \\
	&= \frac{1}{\sigma_L} \ln\left( \frac{C_\ell}{C_\ell(X)} \right)  +   \epsilon_\ell\;,  \\
	&= 	\frac{1}{\sigma_L} \ln\left( 1 + \frac{R_\ell(X)}{C_\ell(X)} \right) + \epsilon_\ell \;,
\end{align}
where $\epsilon_\ell$ is a Gaussian noise with zero mean, introduced in Equation \ref{eq:VST_Cl}. Assuming that the residual $R_\ell(X)$ is small compared to $C_\ell(X)$, one can linearise the above equation, to a good approximation, as
\begin{equation}
	\mathcal{T}(\widetilde{C}_\ell) - \frac{\ln( C_\ell(X) )}{\sigma_L}  \simeq \frac{1}{\sigma_L C_\ell(X)} R_\ell(X) + \epsilon_\ell \;,
\end{equation}
and
\begin{equation}
R_\ell(X) \simeq C_\ell(X) \sigma_L \left( \mathcal{T}(\widetilde{C}_\ell) - \frac{\ln( C_\ell(X) )}{\sigma_L}\right) - C_\ell(X) \sigma_L  \epsilon_\ell \;.
\end{equation}
In this expression, the variance of the noise, i.e. the second term in the above equation, depends on the current estimate $C_\ell(X)$. As we need to estimate the variance of the noise propagated to the wavelet coefficients using Monte-Carlo simulations, it would be too expensive to estimate this every time $C_\ell(X)$ changes. Therefore, we opted for an additional approximation and replace the term $C_\ell(X) \sigma_L$ by $C_\ell(X^0) \sigma_L$ where $X^0$ is now a fixed fiducial power spectrum which can be the initial guess of the solution. We can now introduce the estimator $\overline{R}_\ell(X)$ for $R_\ell(X)$ defined as:
\begin{equation}
	\overline{R}_\ell(X) \equiv C_\ell(X^0) \sigma_L \left( \mathcal{T}(\widetilde{C}_\ell) - \frac{\ln( C_\ell(X) )}{\sigma_L}\right)\;,
\end{equation}
which leads to:
\begin{equation}
	\overline{R}_\ell(X) \simeq  \frac{C_\ell(X^0)}{C_\ell(X)} R_\ell(X) + C_\ell(X^0) \sigma_L  \epsilon_\ell \;.
	\label{eq:add-noise}
\end{equation}
Unless $C_\ell(X^0)=C_\ell(X)$ in the first term, this estimator yields a biased estimate of the amplitude of $R_\ell(X)$. However, it still verifies $\overline{R}_\ell(P_k^{\mathrm{th}}) = 0$ and unless the estimated solution $X$ deviates significantly from $X^0$, the ratio $C_\ell(X^0)/C_\ell(X)$ remains limited to within a few percents. Furthermore, the fiducial power spectrum $X^0$ can be reset several times to the current estimated $X$ as the algorithm converges towards a solution, therefore removing any potential multiplicative bias on the residuals once the algorithm has converged. On the other hand, the noise on the estimator $\overline{R}_\ell(X)$ now has a fixed variance independent of the current estimate of the solution $X$. Replacing this estimator in the data fidelity term of Equation~\ref{eq:bp-lagragian} eliminates the unknown true anisotropy power spectrum from the data fidelity term. 

We furthermore modify the sparsity constraint by applying a weight for each wavelet coefficient thus turning the parameter $\lambda$ in Equation~\ref{eq:bp-lagragian} into $K \lambda_i$, where $i$ is the coefficient index in the wavelet domain. In section \ref{sec:params}, a specific choice of the $\lambda_i$ will allow us to use a single regularisation parameter $K$ to handle the non stationary and correlated noise on the estimator $\overline{R}_\ell$ in a way that translates into a significance level threshold for the detection of features.
The optimisation problem solved by PRISM can now be formulated as:
\begin{equation}
\min\limits_X \frac{1}{2} \parallel \frac{1}{C_\ell(X^0) \sigma_L} \overline{R}_\ell(X) \parallel_2^2 + K \sum_i \lambda_i \parallel [ \mathbf{\Phi}^t X ]_i \parallel_0\;,
\label{eq:bp-lagragian-regularised}
\end{equation}
where the pre-factor ${1}/{C_\ell(X^0) \sigma_L}$ weights the $\ell_2$ data fidelity term according to the variance of the noise on the estimator $\overline{R}_\ell$.

\subsection{The PRISM algorithm}

The $\ell_0$ optimisation problem stated in Equation~\eqref{eq:bp-lagragian-regularised} cannot be solved directly. However, the solution can be estimated by solving a sequence of relaxed problems using the re-weighted $\ell_1$ minimisation technique \citet{Candes2007}. This technique amounts to solving a sequence of weighted $\ell_1$ problems of the form:
\begin{equation}
\min\limits_X \frac{1}{2} \parallel \frac{1}{C_\ell(X^0) \sigma_L} \overline{R}_\ell(X) \parallel_2^2 + K \sum_i \lambda_i | [ \mathbf{W} \mathbf{\Phi}^t  X ]_i  |\;,
\label{eq:reweighted-bp-lagragian}
\end{equation}
where $\mathbf{W}$ is a diagonal matrix applying a different weight for each wavelet coefficient. This relaxed problem is now tractable and the solution of the original problem \eqref{eq:bp-lagragian-regularised} can be approximated using the iterative algorithm presented in \citet{Candes2007} to perform the reweigted analysis-based $\ell_1$ recovery:
\begin{enumerate}
	\item Set $j=0$, for each element of the weighting matrix $\mathbf{W}$ set $w_i^j = 1$. Set the first guess $X^0$ by fitting a pure scale invariant primordial power spectrum to the data $\widetilde{C}_\ell$. 
	
	\item Solve the weighted $\ell_1$ problem \eqref{eq:reweighted-bp-lagragian} yielding a solution $X^j$.
	
	\item Compute $\alpha_i^j = \mathbf{\Phi} X^j$ and update the weights according to:
			\begin{equation}
				w_{i}^{j+1} = \left\lbrace 
				\begin{matrix} \frac{1}{|\alpha_i^j|/K\sigma_i} &\quad \mbox{ if } |\alpha_i^j| \geq K \lambda_i \\
						1  &\quad \mbox{ if } |\alpha_i^j| < K \lambda_i
				\end{matrix}\right. \;,
			\end{equation}
where $\lambda_i$ is the standard deviation propagated to the wavelet coefficients (see section~\ref{sec:params}) and $K$ is a given significance level.

	\item Terminate on convergence or when reaching the maximum number of iterations, otherwise go to step 2.
\end{enumerate}
In practice, we find that three iterations of this procedure are enough to reach satisfying convergence and de-biasing our results and we see no further improvements by performing additional re-weightings.

To solve the relaxed problem \eqref{eq:reweighted-bp-lagragian} given a weighting matrix $\mathbf{W}$, the popular Iterative Soft Thresholding Algorithm (ISTA) can be used. This proximal forward-backward iterative scheme relies on the following iteration:
\begin{align}
\widetilde{X}^{n+1} &= X^{n} + \mu \mathbf{T}^{t} \mathbf{M}^{t} \frac{1}{(C_\ell(X^0) \sigma_L)^2} \overline{R}_\ell(X^{n})\;, \\
X^{n+1} &= \prox_{K  \mu \parallel \lambda \odot W \Phi^t \cdot \parallel_1 } \left(\widetilde{X}^{n+1} \right) \;,
\label{eq:ISTA}
\end{align}
where $\mu$ is an adapted step size and $\prox_{K  \mu \parallel \lambda \odot W \Phi^t \cdot \parallel_1 }$ is the proximal operator corresponding to the sparsity constraint. The gradient descent step $\mu$ has to verify:
\begin{equation}
 0 < \mu \leq \frac{2}{ \parallel \mathbf{T}^t \mathbf{M}^t (C_\ell(X^0) \sigma_L)^{-2} \mathbf{M} \mathbf{T} \parallel } \;,
\end{equation}
where $\parallel \cdot \parallel$ is the spectral norm of the operator.

In the absence of a closed form expression for the proximal operator, its value can be estimated by solving a nested optimisation problem:
\begin{equation}
	\label{eq:prox}
	\left\lbrace \begin{matrix} \hat{u} = \argmin_{| u_i | \leq K  \mu \lambda_i w_i} \frac{1}{2} \parallel \mathbf{\Phi} u - x \parallel^2_2  \\
	\prox_{K\mu \parallel \lambda \odot W \mathbf{\Phi}^t \cdot \parallel_1 } (x ) = x - \mathbf{\Phi} \hat{u}
	\end{matrix} \right. \;.
\end{equation}
We solve this optimisation problem at each iteration of the algorithm, using the Fast Iterative Soft Thresholding Algorithm (FISTA) \citet{FISTA2009}, a fast variant of ISTA. The details of the algorithm solving this weighted problem are provided in Algorithm \ref{alg1}.

\subsection{Choice of wavelet dictionary and regularisation parameter}
\label{sec:params}

As mentioned in the previous section, the regularisation parameter $K$ can be set according to a desired significance level. Indeed, in Equation~\eqref{eq:prox}, it can be seen that the wavelet coefficients $u_i$ are constrained within a weighted $\ell_1$ ball and correspond to the non significant part of the signal. In order to place the radius of this $\ell_1$ ball according to the expected level of noise for each wavelet coefficient, we propagate the noise on the estimator $\overline{R}_\ell$ from Equation~\eqref{eq:ISTA} through the operator $\mathbf{\Phi} \mathbf{T}^t \mathbf{M}^t (C_\ell(X^0) \sigma_L)^{-2}$ and estimate its variance at each pixel and each wavelet scale. In practice, we estimate this noise level using Monte-Carlo simulations of the noise on $\overline{R}_\ell$. We set each $\lambda_i$ to the resulting variance for each wavelet coefficient. As a result, coefficients below $K \lambda_i$ will be considered as part of the noise and one only need to set a global parameter $K$ to tune the sparsity constraint according to the noise level. In the following section, we have chosen to put this parameter to $K=5$, thus robustly removing noise. 

The choice of wavelet $\mathbf{\Phi}$ will impact the performance of the algorithm. In the following study, we use bi-orthogonal Battle-Lemari\'e wavelets of order 1 with 9 dyadic wavelet scales. This choice of wavelet is rather generic and not specifically tuned to a type of primordial power spectrum. More physically motivated dictionaries could be used to reconstruct a specific type of feature predicted by a given theory.
\begin{algorithm}[!htb]
\caption{Weighted analysis-based $P_k$ sparse recovery}
\label{alg1}
\begin{algorithmic}[1]
\REQUIRE $\quad$ \\
Pseudo power spectrum of the data: $\widetilde{C}_{\ell}$, \\
Instrumental noise power spectrum $N_\ell$, \\
First guess primordial power spectrum $X^0$, \\ 
Sparsity constraint parameter $K$, \\
Weights $w_i$ for each wavelet coefficients.

\bigskip

\STATE Initialise $C_\ell^0 = \mathbf{M} \mathbf{T} X^{0}$.
\STATE Compute variance $\sigma_i$ of noise $\sim \mathcal{N}(0,1)$ propagated to wavelet coefficients through $\mathbf{\Phi} \mathbf{T}^t \mathbf{M}^t (C_\ell(X^0) \sigma_L)^{-2}$ from Monte-Carlo simulations.
\bigskip
\FOR{$n=0$ to $N_{\max}-1$}
\STATE $\overline{R}^n_\ell = C_\ell^0 \sigma_L \left( \mathcal{T}(\widetilde{C}_\ell) - \frac{\ln( \mathbf{M} \mathbf{T} X^n + N_\ell )}{\sigma_L} \right)$
\STATE $\widetilde{X}^{n+1} = X^n + \mu  \mathbf{T}^{t} \mathbf{M}^{t} (C_\ell(X^0) \sigma_L)^{-2} \overline{R}^n_\ell$
\bigskip
\STATE \underline{\text{Computing $\prox_{\lambda \mu \parallel \mathbf{W \Phi}^t \cdot \parallel_1 }\;:$}}
\STATE Initialise $u_1 = y_0 = \mathbf{\Phi}^t \overline{X}^{n+1}$, $t_1=1$.
\FOR{$k=1$ to $K_{\max}-1$}
\STATE $\overline{u}_k = u_k + \mu^\prime \mathbf{\Phi} \left(\overline{X}^{n+1} - \mathbf{\Phi}^t u_k \right)$
\STATE $y_k = \overline{u}_k - \mathrm{ST}_{\mu w_i K \sigma_i} \left( \overline{u}_k \right)$
\STATE $t_{k+1} = (1 + \sqrt{1 + 4 t_{k}^2})/2$
\STATE $u_{k+1} = y_k + \frac{t_k -1 }{t_{k+1}} (y_k - y_{k-1})$
\ENDFOR 
\bigskip
\STATE \underline{\text{Update of the reconstruction:}}
\STATE $X^{n+1} = \widetilde{X}^{n+1} - \mathbf{\Phi} u_{K_{\max}}$
\ENDFOR
\STATE {\bf Return:} The reconstructed primordial power spectrum $P_k = X^{N_{max}}$.
\end{algorithmic}
\end{algorithm}

\section{Results}
\label{sec:results}

\subsection{Numerical simulations}

To assess the performance of our non-linear algorithm we perform a series of reconstructions for three different types of primordial power spectra: a near scale-invariant spectrum with $n_s=0.972$ \citep{wmap9}, a spectrum with a small running of the spectral index with $n_s=0.972$ and $\alpha_s=-0.017$ \citep{SPT_cos} and a spectrum with $n_s=0.972$ with a compensated feature around $k=0.03$ Mpc$^{-1}$. The first two simple models are the most favoured by the current data and the spectrum with the feature (investigated in other works, see \citep{Nicholson:2009pi}) is only used to demonstrate the ability of the algorithm to detect and reconstruct isolated features. In all cases, the cosmological parameters responsible for the evolution of the Universe in the radiation transfer function are kept the same and according to the WMAP 9-year parameters \citep{wmap9}; $\Omega_bh^2=0.02264$, $\Omega_ch^2 = 0.1138$, $\Omega_\Lambda=0.721$, $\tau=0.089$. 

For a thorough comparison of our simulations to the WMAP 9-year data we perform the Monte-Carlo simulations at the level of the WMAP five frequency channels, taking into account the propagation of the instrumental noise through the component separation and masking steps. For each of the three test primordial spectra we produce a set of 2000 pseudo power spectra $\widetilde{C}_{\ell}$ by processing the simulated channels through the LGMCA component separation pipeline \citep{Bobin2013} before computing the empirical power spectrum of the masked maps.
In detail, the simulations are produced using the following steps:
\begin{itemize}
	\item \textbf{Frequency channels}: We simulate CMB maps at the five WMAP channels at frequencies 23, 33, 41, 61 and 94 GHz. The frequency dependant beams are perfectly isotropic PSFs and their profiles have been obtained as the mean value of the beam transfer functions at each frequency as provided by the WMAP consortium (9 year version).
	\item \textbf{Instrumental noise}: Noise maps for each channel have been generated as  Gaussian realisations of pixel variance maps obtained by combining the nine 1-year full resolution hit maps as provided by the WMAP consortium.
	\item \textbf{Cosmic microwave background}: Gaussian realisations of the CMB are computed from the three power spectra $C_\ell^{\mathrm{th}}$, which were obtained by applying the radiation transfer function $\mathbf{T}$ to each of the three test primordial power spectra. The transfer function is computed using CLASS\footnote{\url{http://class-code.net/}} \citep{2011Blas} according to the best-fit WMAP 9-year cosmology. The CMB signal for each channel is then obtained by applying the corresponding beam to the simulated CMB map as well as the HEALPix window for $nside$ of 1024.
	\item \textbf{LGMCA Component Separation}: Full sky 15 arcmin resolution maps are obtained by applying LGMCA, with the precomputed set of parameters \citep{Bobin2013}, to the five simulated channels for CMB and noise. Noisy full sky maps are obtained by adding the resulting signal and noise maps.
	\item \textbf{Masking}: Final maps are obtained by applying the WMAP mask kq85 mask with $f_{sky} = 0.75$.
\end{itemize}

\begin{figure}[t]
\includegraphics[width=\columnwidth]{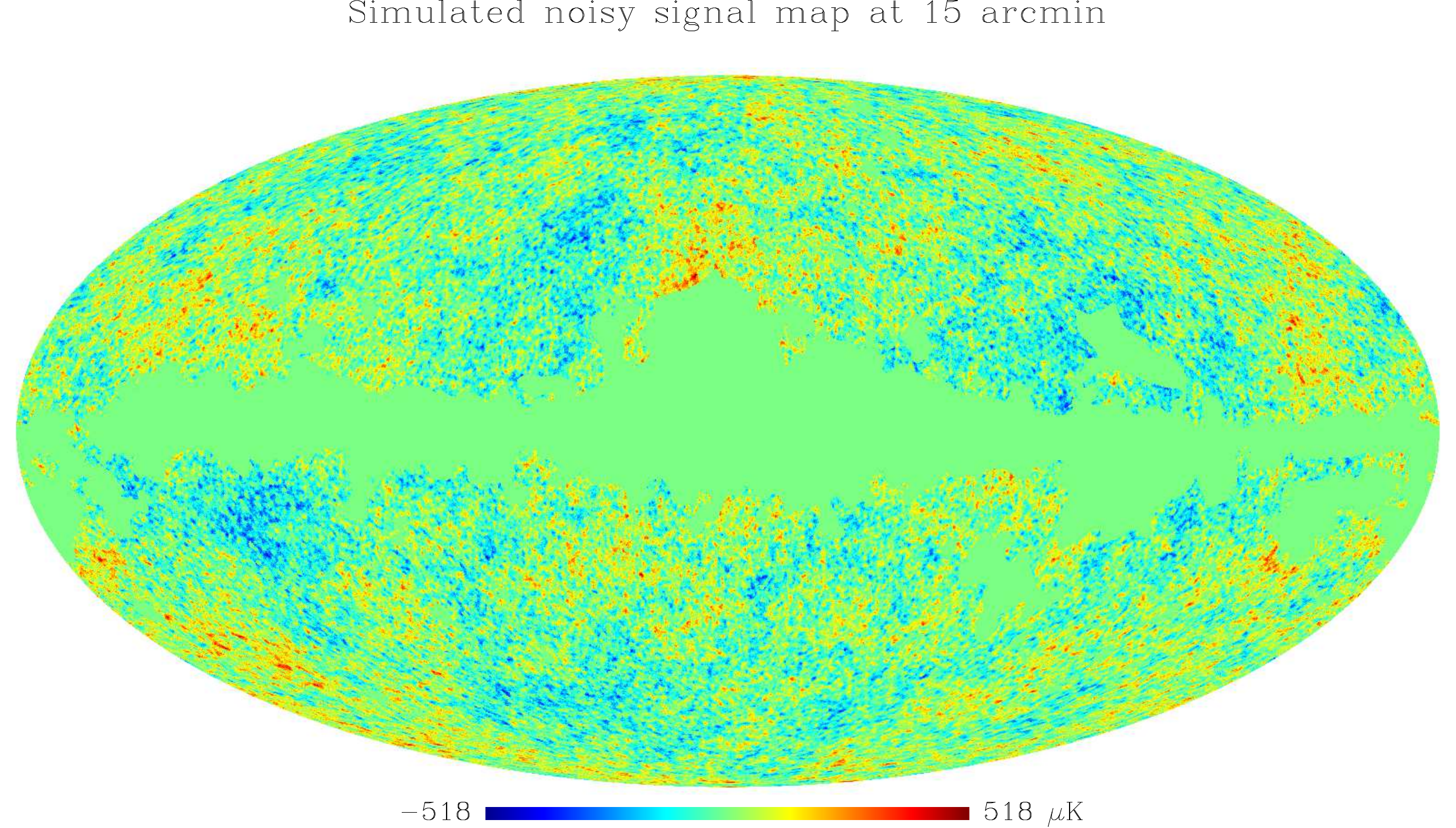}
\caption{A simulated noisy CMB map at 15 arcmin resolution obtained from LGMCA and masked with the WMAP kq85 mask. The noise level corresponds to the WMAP 9-year data. This map was generated from a CMB power spectrum for a primordial spectrum with $n_s=0.972$ and $\alpha_s=0$.}
\label{fig:SimulatedMap}
\end{figure}

\begin{figure}[t]
\includegraphics[width=1.1\columnwidth]{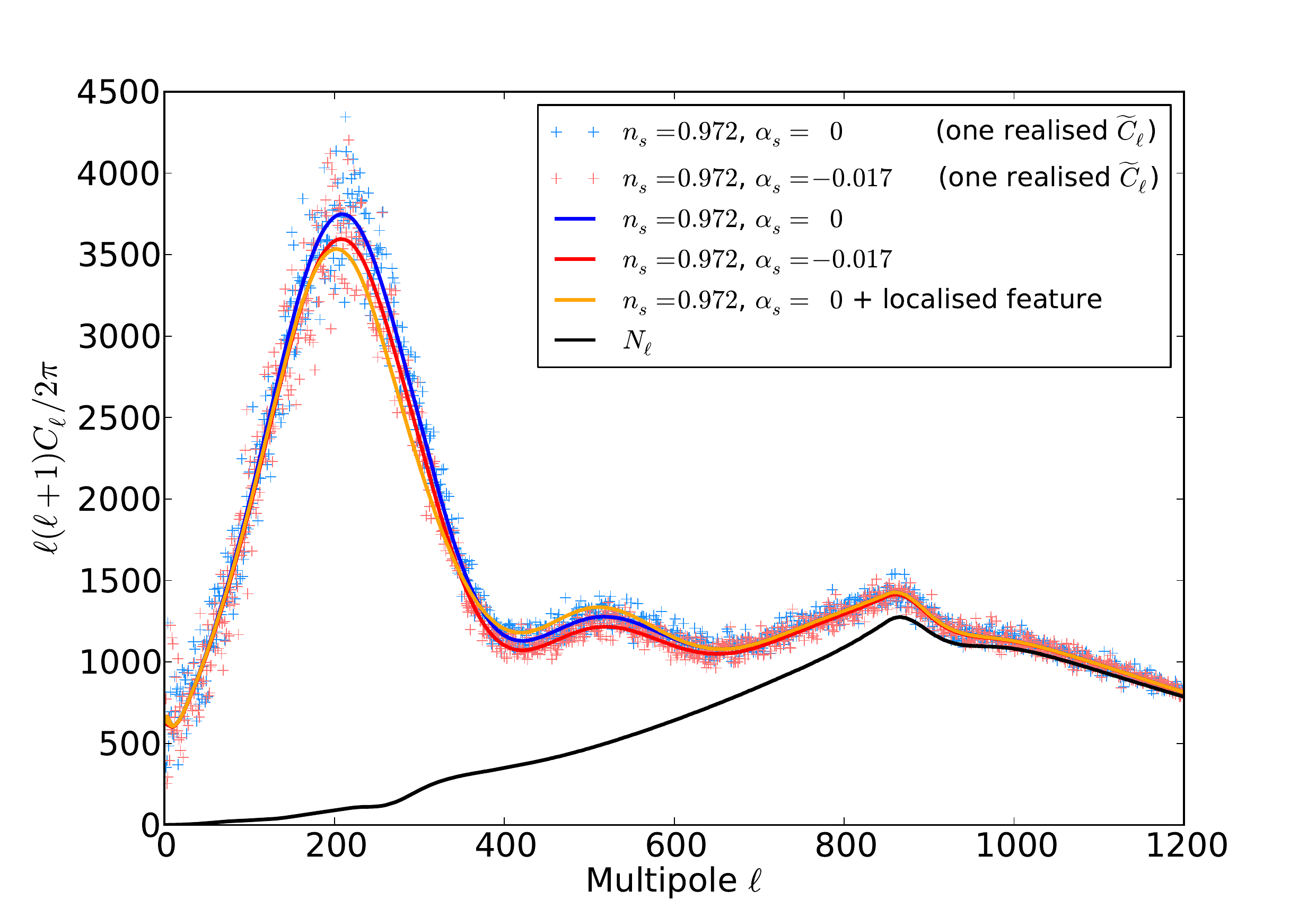}
\caption{CMB pseudo power spectra for the three types of primordial power spectra. The blue solid line shows the pseudo spectrum based on a primordial spectrum with $n_s=0.972$ and $\alpha_s=0$. The light blue crosses show one simulation of this spectrum, computed from the map in Figure~\ref{fig:SimulatedMap}. The red line shows the pseudo spectrum for a primordial spectrum with $n_s=0.972$ and $\alpha_s=-0.017$ and the orange line corresponds to power spectrum with a localised feature at $k = 0.03$ Mpc$^{-1}$. These spectra include the effects of the mask, the 15 arcmin beam, the HEALPix window for $nside$ of 1024 and the instrumental noise power spectrum, which is shown in solid black line.
\label{fig:Clall_simCl}}
\end{figure}

The pseudo power spectra are obtained by applying the empirical power spectrum estimator to the simulated maps. The noise power spectrum $N_\ell$ is estimated by averaging the 2000 pseudo spectra of masked noise maps. Figure~\ref{fig:SimulatedMap} shows an example of a masked noisy CMB map obtained from our simulation process. Figure~\ref{fig:Clall_simCl} shows the pseudo power spectra for the three test primordial spectra as well as the instrumental noise power spectrum estimated from the simulations. The light blue crosses show one realisation of the pseudo power spectrum for the near scale-invariant primordial power spectrum and the pink crosses show the one with a small running. As can be seen, the three different CMB spectra lie well within each others noise band and on large and small scales they become almost indistinguishable. Hence to accurately reconstruct the three underlying primordial power spectra from these CMB spectra, a very good handle on both the instrumental noise and the sample variance is required.

\subsection{Reconstructions of primordial power spectra}

To apply PRISM to the simulated data, we build a transfer function $\mathbf{T}^\prime$ adapted to the simulations so that it includes the effects of the 15 arcmin beam from LGMCA and the HEALPix window of $nside=1024$. Using the same radiation transfer function $\mathbf{T}$ as computed for the simulations, the resulting transfer matrix $\mathbf{T}^\prime$ can be written as:
\begin{equation}
   \mathbf{T}^\prime = \mathbf{Q} \mathbf{T} b_\ell^2 h_\ell^2 \;,
   \label{eq:FinalT}
\end{equation}
where $b_\ell^2$ and $h_\ell^2$ for the beam and the HEALPix window respectively and $\mathbf{Q}$ is an operator performing a linear interpolation from the linear sampling in $k$ of the CLASS transfer function $\mathbf{T}$ to a logarithmic scale using 838 points in the range $k\sim10^{-4} - 0.15$ Mpc$^{-1}$. We also compute the MASTER coupling matrix $\mathbf{M}_{{\textrm{kq}}85}$ corresponding to the kq85 high-resolution temperature analysis mask used in the simulations. 

We now have all the ingredients necessary in our algorithm: $\mathbf{M}_{{\textrm{kq}}85}$, $\mathbf{T}^\prime$ and $\mathbf{\Phi}$, which we use to construct our algorithm and apply it to the $3 \times 2000$ simulated pseudo power spectra. We use the same set of hyper parameters in PRISM for three types of primordial spectra: a $K\sigma$ significance level for the sparsity constraint with $K=5$, 3 reweightings, and $N_{max}=400$ iterations per reweighting.
\begin{figure*}
\begin{subfigure}[b]{0.52\textwidth}
                \includegraphics[width=\textwidth]{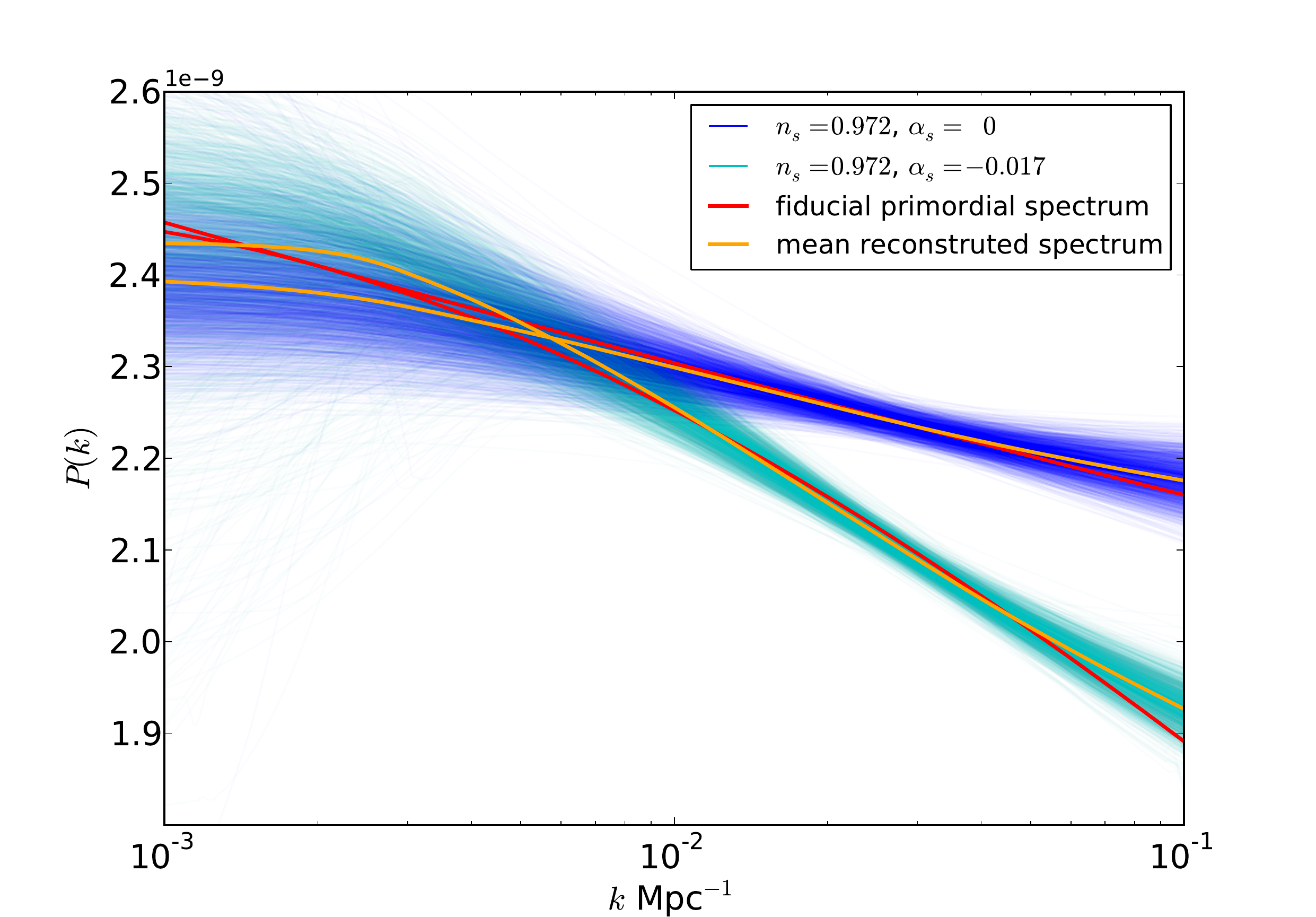}
                \caption{Reconstructed primordial power spectra}
                \label{fig:linrunpk}
        \end{subfigure}%
        ~ 
        \begin{subfigure}[b]{0.52\textwidth}
                \includegraphics[width=\textwidth]{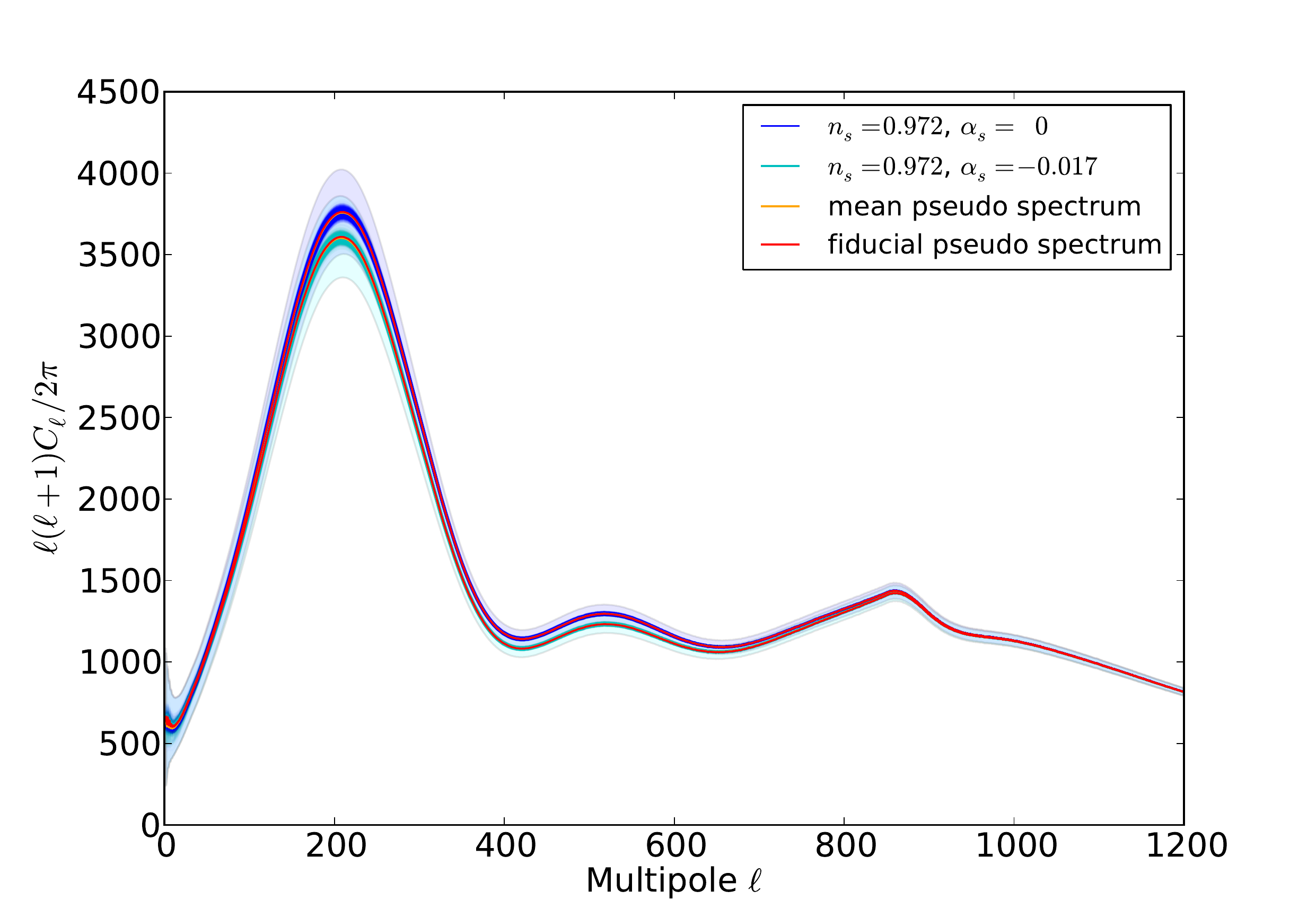}
                \caption{Corresponding CMB pseudo power spectra}
                \label{fig:linruncl}
        \end{subfigure}
\caption{Reconstructions for the primordial power spectra and their corresponding CMB pseudo spectra are shown. In blue we show the 2000 reconstructed spectra with $n_s=0.972$ and $\alpha_s=0$ and in cyan the reconstruction for $n_s=0.972$ and $\alpha_s=-0.017$. In both cases the mean of reconstructions is shown in orange and the fiducial input spectrum is shown in red. As can be seen, for $k>0.015 \; \mathrm{Mpc}^{-1}$ PRISM can reconstruct the primordial power spectra with such accuracy that the two are easily distinguishable, despite their very similar forms in $C_\ell$ space: the shaded regions in the right hand plot correspond to the one-sigma sample (cosmic) variance, which demonstrates the similarity of the two types of CMB spectra. The quality of the reconstruction can also be seen in the reconstructed angular power spectra which are extremely close to the theory and well within the one-sigma sample variance intervals.
\label{fig:Pkall_linrun}}
\end{figure*}

In Figure~\ref{fig:linrunpk} we show the reconstructed primordial spectra in the range $k\sim0.001-0.10 \mathrm{\;Mpc}^{-1}$. The blue lines show the $2000$ reconstructed spectra for the spectrum with $n_s=0.972$ and $\alpha_s=0.0$ and the cyan lines show the reconstructions for the spectrum with $n_s=0.972$ and $\alpha_s=-0.017$. In each case, the orange line is the mean of the reconstructions and the red line is the fiducial one.

The reconstruction of the primordial power spectrum is limited by different effects on different scales. On very large scales, there are fundamental physical limitations placed on the recovery of the primordial power spectrum by both the cosmic variance and the more severe geometrical projection of the modes. The physical limitations in the radiation transfer function places an inherent limitation at large scales meaning the primordial power spectrum cannot be fully recovered on these scales, even in a perfect CMB measurement. On the other hand, on small scales we are limited by the instrumental noise. This leaves us with a window through which we can recover the primordial power spectrum with a good accuracy. Nevertheless, as can be seen, for $k>0.015 \;\mathrm{Mpc}^{-1}$ the PRISM algorithm can reconstruct the primordial power spectrum to a great accuracy and easily distinguish between the two types of spectra. 

Figure~\ref{fig:linruncl} shows the 2000 CMB spectra obtained from the reconstructed primordial power spectra of each type. The blue lines show the CMB power spectra obtained from the near scale-invariant primordial spectra and the cyan lines show the Cones for the primordial spectrum with a running. In each case, the orange line shows the mean of the reconstructions and the red line shows the fiducial one. Comparing these CMB spectra to the input simulated ones, shown in Figure~\ref{fig:Clall_simCl}, shows the great performance of the PRISM algorithm.

\begin{figure}
\includegraphics[width=1.1\columnwidth]{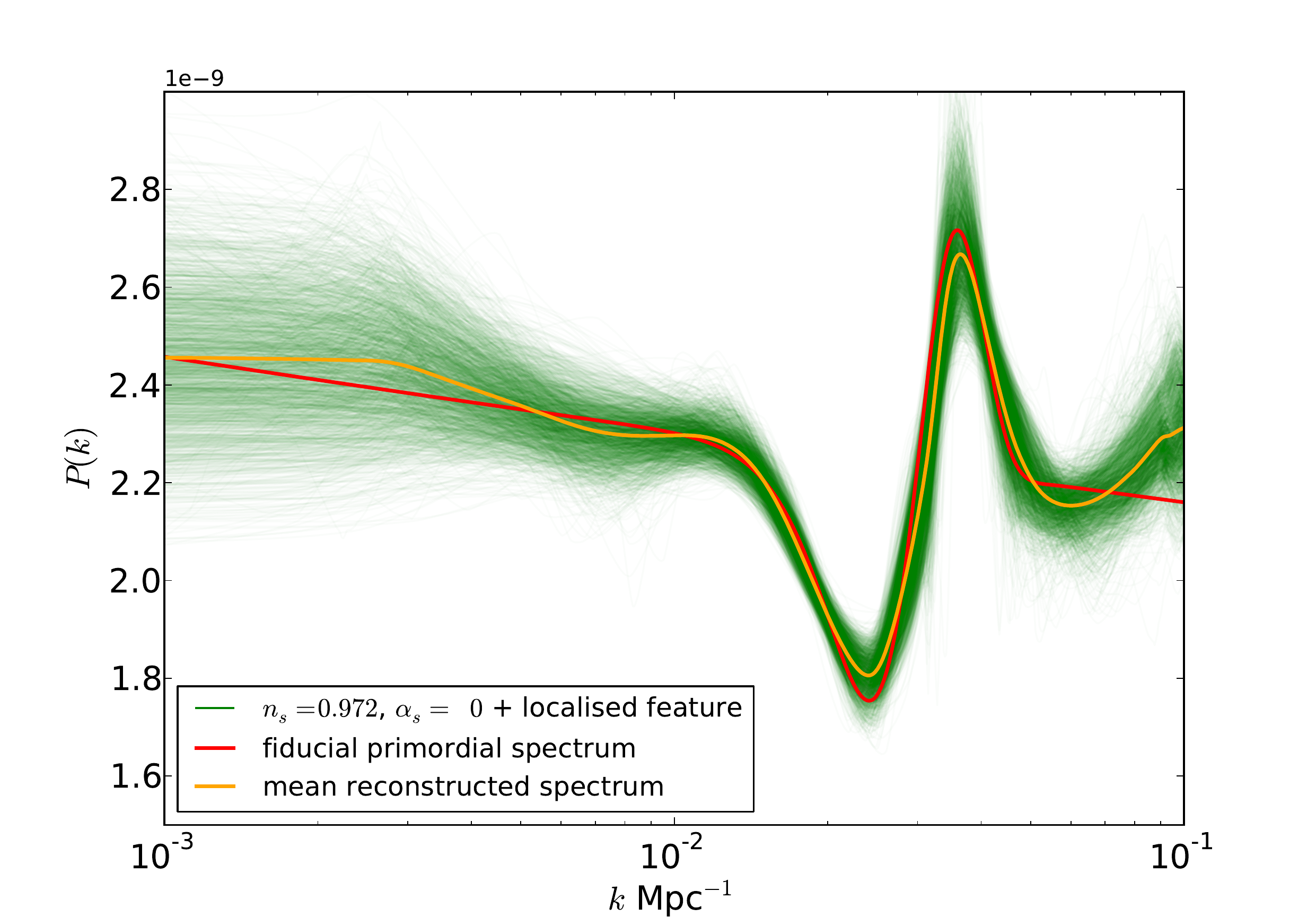}
\caption{Reconstruction of the primordial power spectrum with $n_s=0.972$, $\alpha_s=0.0$ and an additional feature around $k=0.03$ Mpc$^{-1}$ is shown in green. The 2000 reconstructions are superimposed with their mean shown in orange. The fiducial input spectrum is shown in red. As can be seen, PRISM is able to recover both the position and the amplitude of the feature with great accuracy.\label{fig:bmppk}}
\end{figure}

Figure~\ref{fig:bmppk} shows the performance of PRISM in reconstructing a localised feature in the primordial power spectrum. The green lines show the 2000 individual reconstructions, the orange solid line shows the mean of the reconstructions and the fiducial spectrum is shown in red. As can be seen, both the position and the amplitude of the feature can be recovered with great accuracy. 

\subsection{Reconstruction from WMAP 9-year CMB spectrum}

\begin{figure*}
\begin{subfigure}[b]{0.52\textwidth}
                \includegraphics[width=\textwidth]{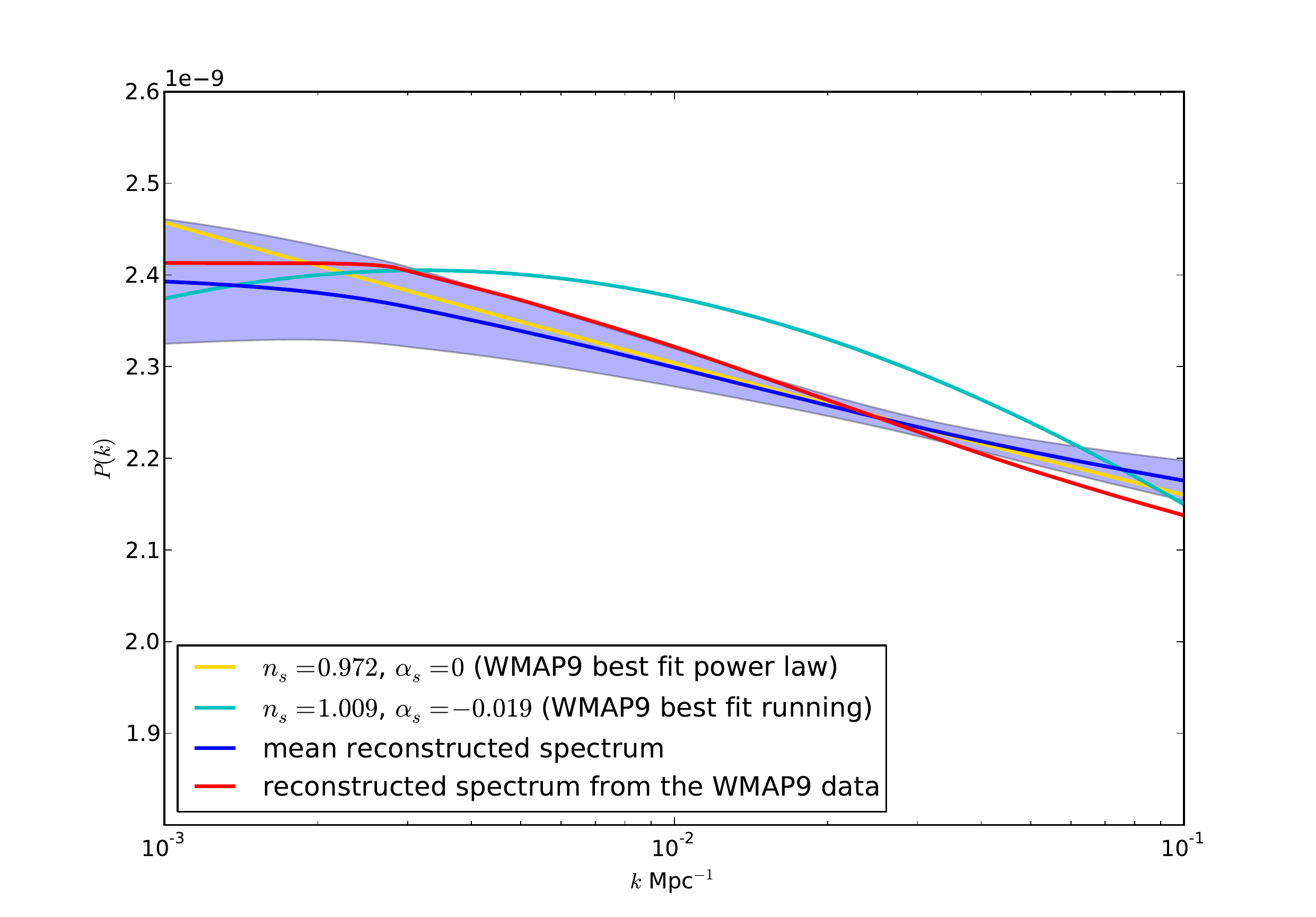}
                \caption{Reconstructed primordial spectrum from WMAP 9-year data}
                \label{fig:wmappk}
        \end{subfigure}%
        ~ 
        \begin{subfigure}[b]{0.52\textwidth}
                \includegraphics[width=\textwidth]{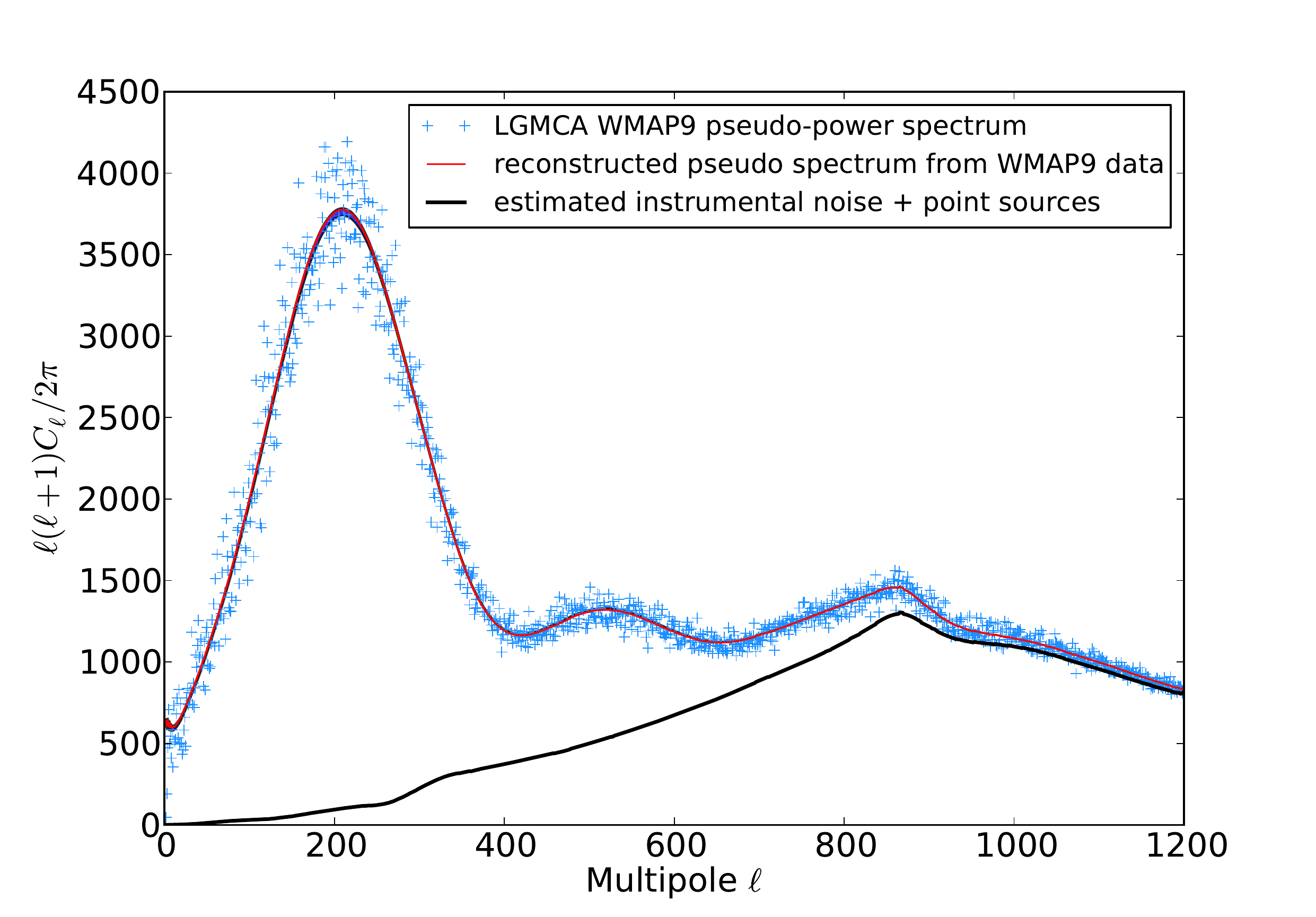}
                \caption{Corresponding CMB pseudo power spectrum}
                \label{fig:wmapcl}
        \end{subfigure}
\caption{Reconstruction of the primordial power spectrum from the LGMCA WMAP 9-year data and its corresponding pseudo spectrum are shown in red. For comparison, we also show the mean of the reconstruction for $n_s=0.972$ and $\alpha_s = 0$ in solid dark blue line with the one-sigma interval around the mean shown as a shaded blue region. The WMAP 9-year fiducial primordial power spectrum with $n_s=0.972$ and $\alpha_s = 0$ is shown in yellow and in cyan we show the best fit primordial power spectrum with a running from WMAP 9-year data with $n_s=1.009$ and $\alpha_s = -0.019$.
On the right, we plot the LGMCA WMAP 9-year pseudo power spectrum (blue crosses) and the estimated instrumental noise power spectrum including the point sources power spectrum is shown (black solid line). The very small blue region corresponds to the one-sigma interval around the mean reconstructed spectrum (i.e. blue region on the left plot).
As can be seen, we do not detect a significant deviation of the WMAP 9-year data from the best fit near scale-invariant spectrum.
\label{fig:Pkall_linrun}}
\end{figure*}

In the WMAP 9-year analysis \citep{wmap9}, the cosmological parameters in the radiation transfer function are fitted along with $n_s$ and $A_s$, hence a power law form for the primordial power spectrum is assumed. This means the transfer function computed using these best fit parameters will always allow a power law primordial power spectrum to fit the observed data. However, reconstructing a free form primordial power spectrum from the data, assuming the fiducial transfer function, allows us to test this null hypothesis by looking for significant deviations between the reconstructed spectrum from data and the simulations. 

The WMAP 9-year data is processed using LGMCA as described in \citet{Bobin2013}, which is the same pipeline used to produce the simulations. As mentioned previously, a good handle on the noise power spectrum is critical in order to yield an unbiased reconstruction of the primordial power spectrum. We estimate the noise power spectrum from the WMAP 9-year data by subtracting the cross-power spectrum from the auto-power spectrum and applying a denoising, using the TOUSI algorithm. To account for the effect of point sources, which were not accounted for in the simulations, we add an estimate of the point sources power spectrum, computed from 100 simulations, to the estimated noise power spectrum. Figure~\ref{fig:wmapcl} shows the pseudo-power spectrum computed from the LGMCA WMAP 9-year map (blue crosses) and the estimated instrumental noise power spectrum (black solid line). Note that in theory, the noise power spectrum could be computed from simulations. However, after comparing our estimated noise power spectrum from the 2000 simulations to the actual noise power spectrum in the WMAP 9-year data we found a small bias that we could not account for in the simulations. Hence we opted for using the data itself to estimate the noise power spectrum.

We apply PRISM, with the same hyper parameters as in the simulations, to the WMAP 9-year LGMCA CMB pseudo power spectrum. The reconstructed primordial power spectrum is shown in red in Figure~\ref{fig:wmappk}. In this figure, we overlay the one-sigma interval around the mean of reconstructed primordial near scale-invariant spectrum, obtained from the simulations. The best fit power law power spectrum from WMAP 9-year with $n_s=0.972$ and $\alpha_s = 0$ is shown in yellow while the best fit power spectrum with a running from WMAP 9-year with $n_s=1.009$ and $\alpha_s = -0.019$ is shown in cyan \citep{wmap9}. As can be seen, the reconstructed power spectrum from data does not exhibit a significant deviation from the best fit near scale-invariant spectrum. The small departure from the one-sigma interval at small scales is not significant, especially since our simulations did not thoroughly take into account additional effects such as a beam uncertainty and point sources. To conclude, we find no significant departure from the WMAP 9-year best fit near scale-invariant spectrum.

\section{Conclusions}
\label{sec:conclusion}

The primordial power spectrum describes the initial perturbations
in the Universe and hence provides an indirect probe of inflation
or other structure-formation mechanisms. The simplest models of inflation are the most favoured by the data and predict a nearly scale-invariant power spectrum with a small running. One way to measure this spectrum is through the windows of the CMB data. The problem, though, is that the singular nature of the radiation transfer function and the joint estimation of the cosmological parameters in the transfer function and the primordial power spectrum, along with the different types of noise sources impose a limit into the full recovery of the primordial spectrum. Therefore, devising a technique which is sensitive enough to detect deviations from scale-invariance is the key to recover an accurate primordial power spectrum.

In this paper we have introduced a new non-parametric technique, coined PRISM, to recover the primordial power spectrum from masked noisy CMB data. This is a sparse recovery method, which uses the sparsity of the primordial power spectrum as well as an adapted modelling for the noise of the CMB power spectrum. This algorithm assumes no prior shape for the primordial spectrum and does not require a coarse binning of the power spectrum, making it sensitive to both global smooth features (e.g., running of the spectral index) as well as local sharp features (e.g., a bump or an oscillatory feature). Another advantage of this method is that, thanks to the clever modelling of the sample variance on the input angular power spectrum, the regularisation parameter can be specified in terms of a signal-to-noise significance level for the detection of features. These advantages make this technique very suitable for investigating different types of departures from scale-invariance in the primordial power spectrum, whether it is the running of the spectral index or some localised sharp features as predicted by some of the inflationary models. 

We have investigated the strength of our proposed algorithm on a set of WMAP 9-year simulated data for three types of primordial power spectrum; a near scale-invariant spectrum, a spectrum with a small running of the spectral index, and a spectrum with a localised feature. We have shown that our algorithm can easily recover the three spectra with an excellent accuracy in the range $k\sim0.001-0.1 \mathrm{\;Mpc}^{-1}$. In addition, the errors in the recovered spectra are small enough that the three types of primordial spectra can easily be distinguished in the range $k\sim0.015-0.1 \mathrm{\;Mpc}^{-1}$. This technique has proved to easily detect small global and localised deviations from a pure scale-invariant power spectrum and is suitable for distinguishing between simple models of the inflation.

Using PRISM, we have reconstructed a primordial power spectrum from the LGMCA WMAP 9-year data and have investigated possible departures from the WMAP 9-year near scale-invariant spectrum. We have not detected any significant deviations from this simple model of the primordial power spectrum. We have demonstrated the feasibility of using PRISM on masked CMB data contaminated by instrumental noise. Better constraints will be obtained in future works by processing Planck data which provides a much lower instrumental noise, thus improving the range of scales we are able to probe with much better accuracy.

To this end, we also acknowledge previous algorithms aimed at reconstructing the primordial power spectrum with no need for binning, most of which have been referenced in this paper. The most recent work is by \citet{HazraEtAl20131}, who use an adapted and improved Richardson-Lucy algorithm, dubbed MRL, to reconstruct the primordial power spectrum. Due to the very high level of the instrumental noise on small scales in the WMAP 9-year data, the MRL algorithm can only take the unbinned CMB spectrum for $\ell<900$. For larger angular scales, $\ell=900-1200$, a binned CMB spectrum is used. In addition, due to the induced artefacts in the reconstructed primordial spectrum, a smoothing step is necessary after the reconstruction is performed. Henceforth, compared to the MRL algorithm, the advantage of our algorithm is twofold. One is the ability to use the unbinned CMB spectrum for the whole multipole range $\ell=2-1200$. This is because of our accurate noise modelling on the CMB power spectrum. In addition, unlike the MRL algorithm, there is no need to smooth the spectrum after the reconstruction as we look for the sparsest solution in our algorithm. This is why PRISM performs significantly better than previous algorithms, including the MRL algorithm.

The developed C++ and IDL codes will be released with the next 
version of iSAP (Interactive Sparse astronomical data Analysis Packages) via the web site\\ \\
{\centerline{\texttt{http://cosmostat.org/isap.html} .}}\\ \\
All results have been obtained using the isap routine \textbf{mrs\_prism} with the following command line:
\begin{verbatim}
pk = mrs_prism(Cl, noise=Nl, TransferMat=Mat)
\end{verbatim}
where \texttt{Cl} contains the observed pseudo-power spectrum of the masked noisy CMB maps, \texttt{Nl} is an estimate of the instrumental noise power spectrum $N_\ell$, and \texttt{Mat} is the input transfer matrix which include the effects of the radiation transfer function, the mask, the beam and the HEALPix window (Equation \ref{eq:FinalT}). The transfer matrix can be been computed using the isap routine \textbf{mrs\_transfer\_matrix} and by default the transfer matrix is derived from the WMAP 9-year best fit cosmology model.

\section*{Acknowledgments}
The authors would like to thank Amir Hajian, Gabriel Rilling and Jeremy Rapin for their useful discussions.
This work is supported by the European Research Council grant SparseAstro (ERC-228261)

\bibliographystyle{aa}
\bibliography{biblio}

\end{document}